\documentclass[conference]{IEEEtran}
\IEEEoverridecommandlockouts
\usepackage{cite}
\usepackage{amsmath,amssymb,amsfonts}
\usepackage{algorithmic}
\usepackage{graphicx}
\usepackage{textcomp}
\usepackage{xcolor}
\usepackage[T1]{fontenc}
\usepackage[utf8]{inputenc}
\usepackage[linesnumbered,ruled,vlined]{algorithm2e}
\def\BibTeX{{\rm B\kern-.05em{\sc i\kern-.025em b}\kern-.08em
    T\kern-.1667em\lower.7ex\hbox{E}\kern-.125emX}}
\begin{document}

\title{Exploration of Hyperledger Besu in Designing Private Blockchain-based Financial Distribution Systems}

\author{
\IEEEauthorblockN{
1\textsuperscript{st} Md. Raisul Hasan Shahrukh,
2\textsuperscript{nd} Md. Tabassinur Rahman,
3\textsuperscript{rd} Nafees Mansoor
}
\IEEEauthorblockA{
\textit{Department of Computer Science \& Engineering} \\
\textit{University of Liberal Arts Bangladesh}\\
Dhaka, Bangladesh \\
1\textsuperscript{st} raisul.hasan.cse@ulab.edu.bd,
2\textsuperscript{nd} tabassenur.rahman.cse@ulab.edu.bd,
3\textsuperscript{rd} nafees@ieee.org}
}

\maketitle

\begin{abstract}
Blockchain, a decentralized technology that provides unrivaled security, transparency, and process validation, is redefining the operational landscape across numerous industries. This article focuses on the development of an innovative consortium blockchain-based financial distribution application. This paper illuminates the transformative role of blockchain technology in a variety of sectors by drawing on a plethora of academic literature and current industry practices. It demonstrates the diverse applications of blockchain, ranging from remittances to lending and investments in finance to data administration in healthcare and supply chain tracking. The paper reveals the design and potential of a consortium blockchain-based application for financial distribution. Utilizing the capabilities of Hyperledger Besu, the application is tailored to improve security, scalability, and interoperability, thereby contributing to a more integrated financial ecosystem. The investigation sheds light on the combination of consortium blockchain’s controlled access and Hyprledger Besu’s comprehensive functionality, proposing a secure, transparent, and efficient financial transaction environment. The investigation serves as a resource for academics, industry professionals, and policymakers alike, highlighting the vast potential of blockchain technology, enabled by platforms such as Hyperledger Besu, in accelerating the evolution of traditional systems toward a more decentralized, secure, and efficient future
\end{abstract}

\begin{IEEEkeywords}
Blockchain Technology, Financial Distribution, Hyperledger Besu, Smart Contracts, Process Automation
\end{IEEEkeywords}

\section{Introduction}
Blockchain technology, with its decentralized, secure, and transparent attributes, is promoting innovation across various sectors \cite{nb1}. Hyperledger Besu, an open source Ethereum client, contributes to this momentum by offering increased versatility and functionality \cite{nb2}. The industry's shift toward blockchain solutions is supported not only by the technology's inherent advantages in security, validation, and transparency \cite{nb3}, but also by platforms such as Hyperledger Besu, which offer tailored solutions for the financial sector and enable the development of sophisticated financial applications \cite{nb4}.

The rapid advancements in the financial sector have brought forth the need for enhanced transparency, security, and efficiency in financial transactions and processes \cite{nb5}. While blockchain technology offers the potential to address these challenges, the practical system, especially through platforms like Hyperledger Besu, remains uncertain \cite{nb4}. Traditional financial operations often suffer from inefficiencies, vulnerability to fraud, and a lack of real-time transparency \cite{nb6}. Consortium blockchain platforms, with their decentralized nature and embedded smart contract functionality, promise a paradigm shift \cite{nb5}. However, the seamless integration of such platforms into the financial distribution industry, ensuring interoperability and scalability while maintaining security, remains a challenge \cite{nb7}. This research seeks to bridge this gap by proposing and developing a consortium blockchain-based system with a custom tailored smart contract that utilizes Hyperledger Besu and mitigates the pain points in the traditional financial distribution systems \cite{nb8}.

The proposed system is a consortium blockchain enabled financial distribution system that is anticipated to substantially contribute to the financial distribution industry. Using the interoperability, scalability, and controlled access of a consortium blockchain in conjunction with the robust functionality of Hyperledger Besu, the system seeks to create a more integrated, secure, and efficient financial ecosystem . Moreover, embedded smart contracts will further automate processes, reducing the need for manual intervention and the likelihood of human error. The design of the proposed system prioritizes both adaptability and resilience, thereby addressing long-standing issues in the financial distribution industry. Not only does this integration facilitate operations, but it also paves the way for innovations in financial service delivery. When considered comprehensively, these advancements position the proposed system as a transformative solution.

This research endeavors to highlight the specific strengths and potentialities of Hyperledger Besu, with a particular emphasis on financial distribution, amidst the growing discourse on blockchain's numerous applications. The most important contribution of this paper is its thorough examination of smart contract development and its seamless integration into a private Hyperledger Besu environment. In doing so, the study highlights the transformative potential of smart contracts in automating processes, reducing manual intervention, and lowering traditional errors such as inefficiencies, vulnerability to fraud, and a lack of real-time transparency, that are essential to the evolving financial sector. Furthermore, by delving into the architectural complexities and operational capabilities of this integration, the research offers researchers, industry practitioners, and policymakers invaluable insights. The overarching objective is to clarify how platforms such as Hyperledger Besu, when utilized skillfully, can accelerate the decentralization, security, and efficacy of future financial systems.
Moreover, the paper aims to illuminate future development directions based on industry trends and the anticipated impact of the proposed system \cite{nb9, nb10}. 

This study investigates the capabilities of Hyperledger Besu, focusing on its effectiveness for financial systems. The second section examines extsisting blockchain solutions, while the third section presents the proposed system along with a diagram. The incorporation the proposed system's of smart contract within the Hyperledger Besu framework is described in Section 4. The paper is concluded in section Section 5, by discussing future prospects and technological potential.

\section{Existing Works}
This section examines and dissects a number of noteworthy studies in the area of blockchain technology. These studies, which cover a range of fields, provide insights into the possible uses and difficulties of blockchain technology. The review's goal is to summarize significant results and provide insight into how blockchain technology is being used in a variety of fields.\newline
A blockchain-based system called MudraChain is suggested by the authors \cite{nb11} as a way to alleviate the inefficiencies in current check clearance procedures in financial institutions. This paper stands out for its addition to the conversation on how blockchain will revolutionize banking and finance. Although it makes a substantial contribution to this discussion, it doesn't address the potential difficulties that may arise from the requirement for the widespread use of blockchain technology. The viability and success of MudraChain depend on this adoption, and a more thorough investigation of viable solutions could strengthen the case. The authors in the research paper \cite{nb12} introduce Fairledger, a fair blockchain protocol for financial organizations. The crucial problem of confidence in decentralized systems is skillfully handled by the writers. They suggest fairness, a solution that is built right into the protocol. Nevertheless, despite the positive qualities of Fairledger, the authors should add more depth to the study by going into potential implementation roadblocks, such as technological constraints and user acceptance, as well as solutions to overcome them.\newline
The InterPlanetary File System (IPFS) and blockchain technologies are used for credential verification \cite{nb13}. The authors \cite{nb13} convincingly demonstrate the effectiveness of their suggested solution, Verifi-Chain. However, given the potential for exponential expansion in credential verification databases, the subject may use a more in-depth examination of scalability difficulties. A useful case study on the implementation of a cross-border transaction system employing a consortium blockchain in Shenzhen, China, is provided in the research paper \cite{nb14}. It emphasizes how blockchain technology could support more safe and effective global financial transactions. However, the research falls short of taking into account the more general regulatory problems present in many jurisdictions. The research findings' global relevance and applicability would be improved by a more thorough treatment of this element. The authors \cite{nb15} suggest a consortia blockchain-based decentralized stock exchange platform. This essay makes an important contribution to the conversation on blockchain's potential advantages in the financial markets. However, a comprehensive analysis of the potential market hazards that such a platform would entail is lacking. A thorough examination of these hazards would offer a more impartial viewpoint and strengthen the plan.\newline
Honestchain, a consortium blockchain created for secure health data sharing, is introduced in the research \cite{nb16}. This research deserves praise for its careful examination of the potential of blockchain for healthcare data management. However, given the importance of patient consent and data access controls in the context of personal health data, it might benefit from more discussion of these concerns. The research paper \cite{nb17} suggests a vehicle-to-grid network privacy security solution based on a consortium blockchain and attribute-based signature. Although this makes a substantial contribution to discussions about transportation security, the research only skims the surface of potential implementation difficulties. Technical difficulties, user acceptance issues, and the requirement for widespread implementation of blockchain in transportation networks may be some of these. Although the authors \cite{nb18} do not explicitly address blockchain, they provide a routing system for CR-VANETs. This offers useful information that might guide upcoming blockchain-based applications in automotive networks. The paper's contribution to the area could be improved with more research into how blockchain can be included in such protocols. The consortium blockchain technology for safe data exchange and storage in vehicle networks is introduced in the research paper \cite{nb19}. Although the report offers insightful information, more research may be done on the system's potential performance and scalability, especially in the context of high-speed vehicle networks where rapid data interchange is essential.\newline
The possibilities for combining routing protocols with blockchain systems could be better explained by the in-depth analysis of UAV routing protocols described by the authors \cite{nb20}. Despite being thorough, the article might yet go further to give a more futuristic viewpoint by examining the synergistic potential between these cutting-edge technologies. A blockchain-based approach to dynamic healthcare data access permission is presented in the research \cite{nb21}. Although the research provides a unique method for managing healthcare data, it might investigate more of the difficulties associated with policy and regulatory compliance. Understanding and managing these complications is essential for the viability of the suggested strategy because this industry is highly regulated.
The authors \cite{nb22} explore the application of blockchain to coal mine safety. Although the article offers a novel viewpoint on blockchain applications in non-financial industries, it should do more to address potential barriers to technology adoption in fields that have historically used less of it.\newline
The research done by the authors \cite{nb23} suggests a blockchain-based architecture for pensions. The article makes a compelling argument for blockchain's application to pension management. It might, however, go into further detail about possible user acceptance issues, which is crucial when introducing new technology in settings where users might not be as tech-savvy. Finally, the authors \cite{nb24} suggest a consortia blockchain-based peer-to-peer file storage system. The authors persuasively suggest that traditional file storage methods could be disrupted by blockchain technology. However, given the unchangeable nature of blockchain data, the article might offer a more detailed investigation of problems relating to data recovery and deletion in a blockchain setting. Collectively, these studies offer a thorough grasp of the numerous applications and difficulties of applying blockchain technology across several industries. But a recurrent topic in numerous publications is the need for further investigation of implementation difficulties, such as user adoption, legal obstacles, and scalability problems.\newline
While the literature provides valuable blockchain applications in various industries, there are persistent implementation-related gaps. Studies on MudraChain \cite{nb11} and Fairledger \cite{nb12} provide innovative financial solutions, but they ignore adoption barriers and technological constraints. Similarly, the papers on Verifi-Chain \cite{nb13} and a cross-border transaction system \cite{nb14} lack an exhaustive discussion of scalability and regulatory issues, respectively. Other research \cite{nb15,nb16,nb17,nb18,nb19,nb20,nb21,nb22,nb23,nb24} focuses on blockchain solutions but ignores market risks, data access controls, and data recovery and deletion. These limitations highlight the need for he proposed system, a consortium blockchain-based platform utilizing Hyperledger Besu and smart contracts, which is intended to comprehensively address these obstacles and contribute to the holistic development of blockchain applications across financial distribution domains.

\section{Proposed system}
This research paper's proposed system represents a revolutionary step toward integrating blockchain technology, specifically Hyperledger Besu, into the financial industry. This system is primarily concerned with implementing a smart contract-based solution for the secure, transparent, and efficient distribution of funds within an organization. The system utilizes Hyperledger Besu's private network capabilities to assure data privacy while maintaining blockchain integrity. 

Moreover, the system meticulously integrates the use of a Solidity-programmed smart contract for precise financial transactions. These include managing recipients, ensuring sufficient funds for transactions, and securely connecting bank account information with recipient addresses. The ultimate goal is to build a robust, secure, and traceable system that will revolutionize financial transactions, thereby fostering accountability and trust within the financial ecosystem.

\subsection{Proposed System's Function Flow}
To enable safe, transparent, and effective financial transactions, the proposed application makes use of blockchain technology, specifically the Hyperledger Besu Ethereum client. The program functions as a controlled mechanism for money distribution inside a private, permissioned blockchain network by utilizing a smart contract written in solidity. An organization that wants to transmit money to a peer starts the application's process flow. 

The organization communicates with the smart contract, which includes the terms and circumstances of transaction including fund sufficiency and recipient authorization. A transaction block is constructed and then broadcast to the network once these requirements are met. Here, Hyperledger Besu enters the picture. The network's Ethereum client that the peers use to verify the transaction block is represented by Hyperledger Besu. The task of validating the transaction block in accordance with the Ethereum protocol and the chosen consensus technique is carried out by the Besu nodes inside the network.
\begin{figure}[htp]
    \centering
    \includegraphics[width=\linewidth]{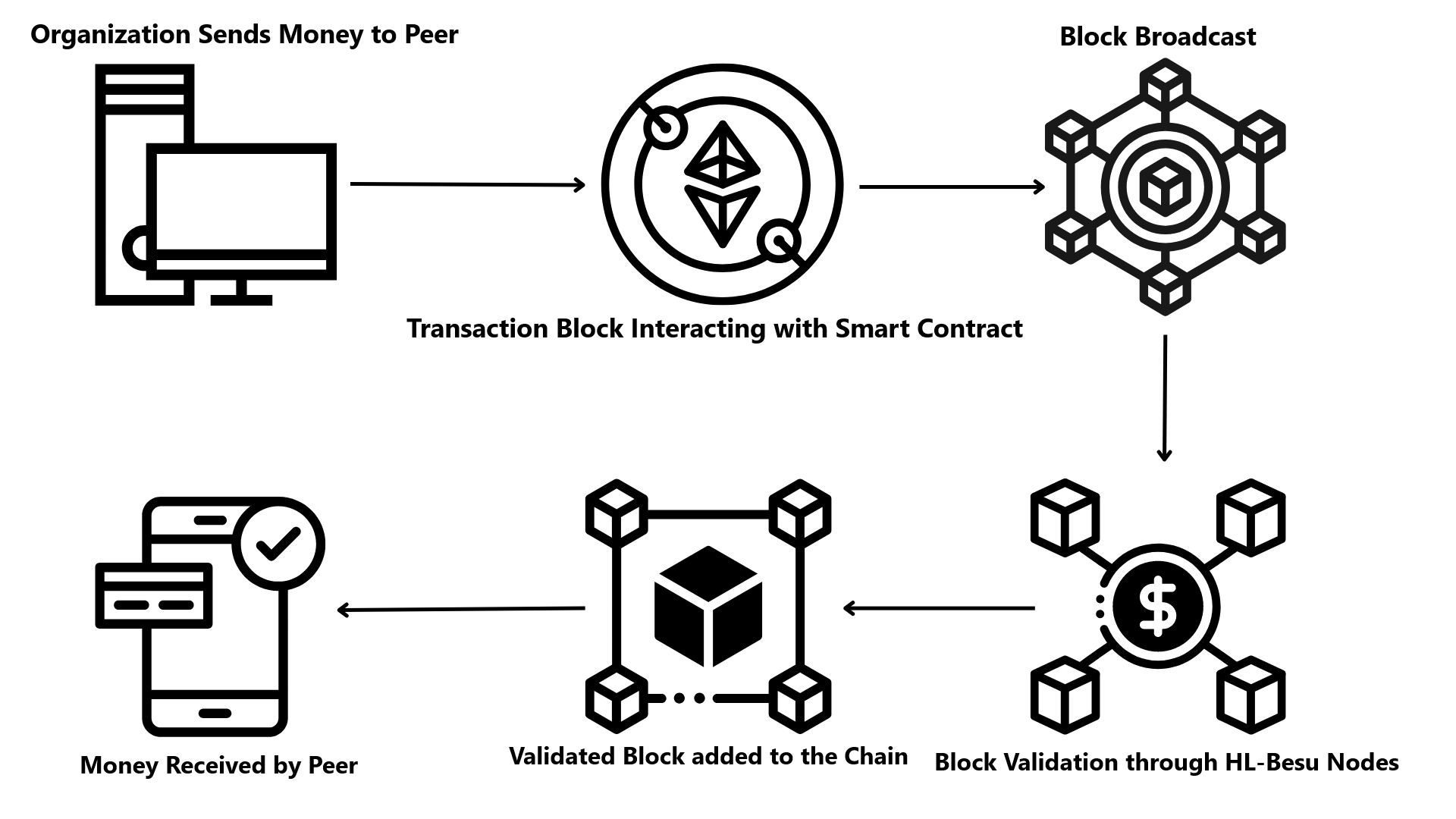}
    \caption{Proposed System's  Block Diagram}
    \label{fig:Proposed System Block Diagram}
\end{figure}
Once the block has been verified, it is put on the blockchain, which serves as a secure record of the transaction. This unchangeable record provides total transaction transparency, boosting the application's security and accountability. 

The peer then receives the funds from the organization after the validated block has been put to the blockchain. The application thereby guarantees an automated, transparent, and immutable financial transaction procedure that is secure, easy, and quick. As a result, this application's function and flow combine the dependability of blockchain technology, notably the Hyperledger Besu Ethereum client, with the accuracy and security of a smart contract to provide a dependable and effective platform for managing and carrying out financial transactions.

\subsection{Structural Diagrams}
The architecture of the proposed system is depicted by two important diagrams; the component and communication diagrams. The component diagram [Fig. 2] illustrates the interconnections between modules such as the User Interface, Blockchain Interface, Hyperledger Besu network, Payment Gateway, and Database. It emphasizes blockchain's function in system operations and transaction security. The communication diagram [Fig. 3] illustrates the dynamic interactions initiated by user actions and facilitated by the smart contract among these entities over time. It clarifies Hyperledger Besu's integral function in transaction validation and propagation. Together, the two diagrams provide a concise representation of the system's architecture and inter-component communication, emphasizing its robustness and operational dynamics.

\subsubsection{Proposed System's Component Diagram}
A more thorough understanding of the financial distribution system is provided in the component diagram[Fig. 2]. It includes the User Interface (UI), Organization Interface, Hyperledger Besu Platform, Smart Contract, and Recipients as its five main parts. The main point of contact for end users or receivers is the user interface. This might be a web app or a mobile app made to make it easier for people to interact with the system. The Hyperledger Besu platform is directly communicated with, enabling users to check their account balances and transactions. The Organization Interface also acts as the point of contact for administrators.
 
\begin{figure}[htp]
    \centering
    \includegraphics[width=\linewidth]{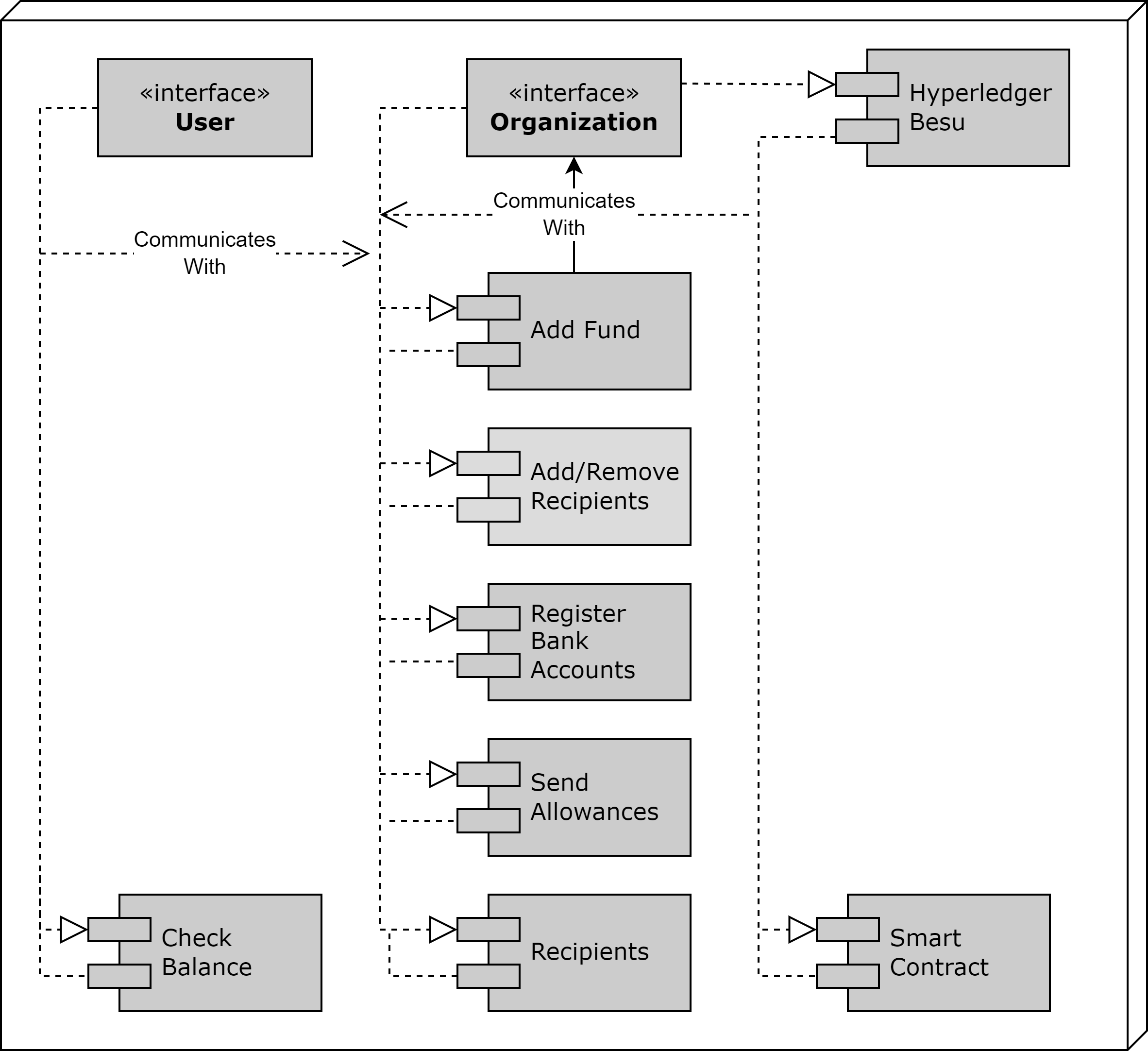}
    \caption{Proposed System's Component Diagram }
    \label{fig:Proposed System's Component Diagram }
\end{figure}

Using this interface, the organiation can administer the smart contract by adding or removing beneficiaries, adding funds, and registering bank accounts, among other operations. Additionally, this interface has direct communication with the Hyperledger Besu platform. The blockchain platform on which the smart contract is implemented is the Hyperledger Besu platform. Between the smart contract and the User and Organization Interfaces, it acts as a bridge, exchanging information.
The business logic of the system, such as recipient management and fund distribution, is encapsulated in the smart contract. The recipients of the funds are represented by the recipients. The User Interface is how they communicate with the system. Finally, the Bank Accounts component represents the system-connected bank accounts of the receivers. Based on the responses from the smart contract, information pertaining to these accounts is displayed on the user interface.

\subsubsection{Proposed System's Communication Diagram}
The communication diagram [Fig. 3] shows how the parts of the system communicate dynamically. It demonstrates how information is sent between the smart contract, the Organization Interface, the User Interface, and the Hyperledger Besu platform. Operations like addRecipient(), removeRecipient(), addFunds(), and registerBankAccount() are used by the Organization Interface to interface with the Hyperledger Besu platform. These operations are sent to the smart contract, which then modifies its state as required. In order to request balance information, the User Interface converses with the Hyperledger Besu platform simultaneously.\newline
The getBalance() call does this, after which the smart contract is contacted. Through the Hyperledger Besu platform, the smart contract obtains the requested balance and transmits it back to the user interface. The smart contract generates an AllowanceSent event whenever money is transferred over the network.
\begin{figure}[htp]
    \centering
    \includegraphics[width=\linewidth]{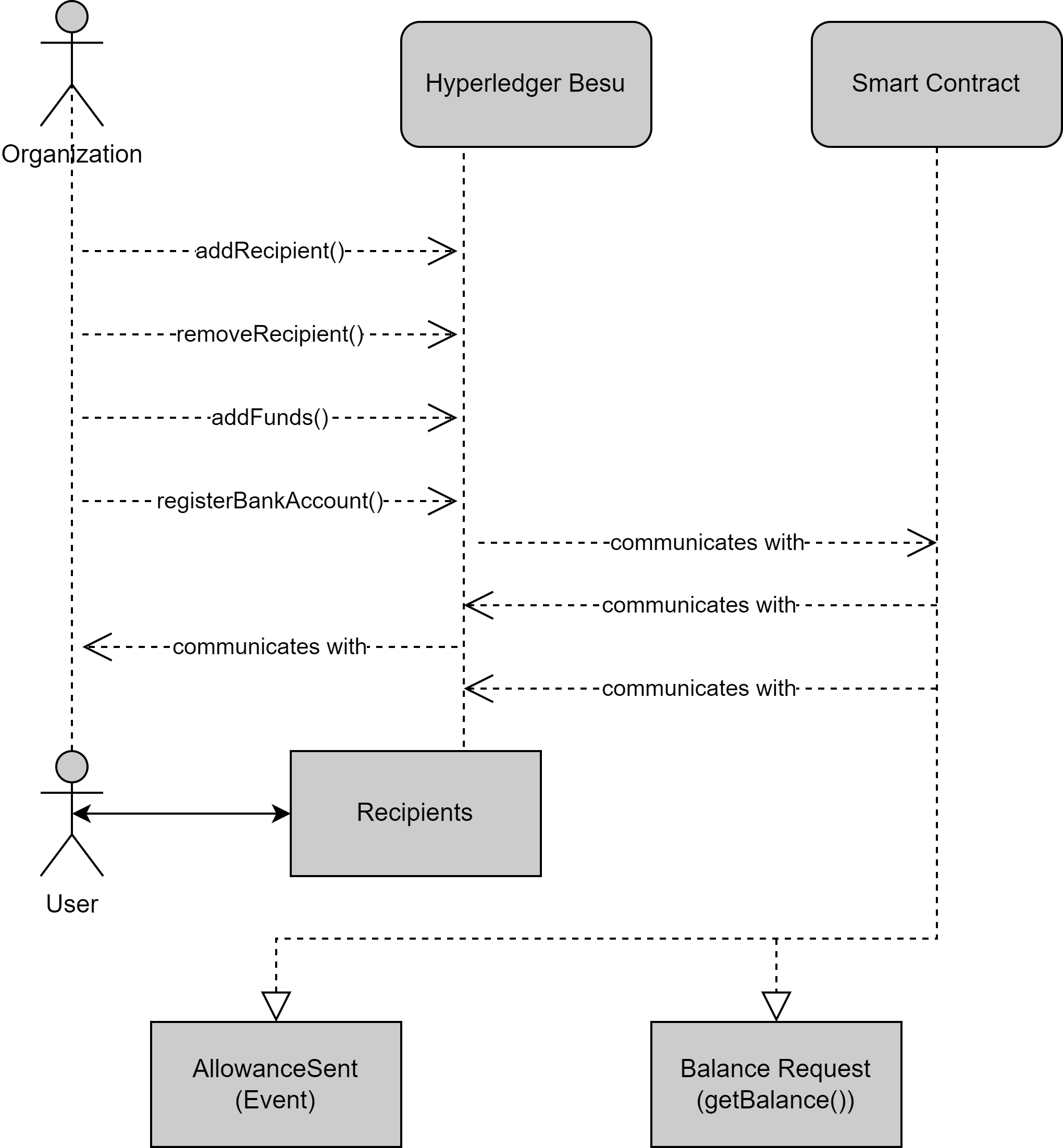}
    \caption{Proposed System's Communication Diagram}
    \label{fig:Proposed System's Communication Diagram}
\end{figure}
The Hyperledger Besu platform transmits this event to the User Interface, informing the user of the transaction. This section examines the proposed system's main operational components. It describes the functionality of the custom-built smart contract that facilitates the secure distribution of financial allowances within a regulated organization. This contract handles recipient management, balance verification, and secure bank account linking, ensuring secure transactions and data integrity. The communication diagram successfully explains the system's operational dynamics in this way by showing how the various parts work together and interact to form an effective and coherent financial distribution system.

\section{Development \& Implementation of Proposed system}
This section describes the process of integrating Hyperledger Besu, a private Ethereum network. The deployment of the smart contract onto this network and subsequent interaction with the system's user interface are described in detail, illuminating the critical role that Hyperledger Besu plays in providing a secure, transparent, and efficient transaction environment.

\subsection{Smart Contract}
Constructed in Solidity, the smart contract functions as a financial disbursement tool within a consortium blockchain, enabling dynamic beneficiary management. Designed to supervise and safeguard financial allocations, the contract ensures transactions comply with stringent verification requirements. It associates recipient addresses with bank details, thereby enhancing data privacy. This system utilizes the capabilities of blockchain to ensure the secure, transparent, and efficient distribution of funds.
\begin{algorithm}
\LinesNotNumbered
\DontPrintSemicolon
\SetAlgoNlRelativeSize{-1}
\SetNlSty{}{}{}
\SetAlgoNlRelativeSize{0}
\SetKwInOut{KwData}{Input}
\SetKwInOut{KwResult}{Output}
\SetKwProg{Fn}{Function}{}{end}
\SetKw{KwRet}{Return}

\KwData{msg.sender: Current function caller}
\KwResult{Constructor(): Initialize the contract with msg.sender as the organization}

organization $\gets$ msg.sender\;

\Fn{AddRecipient(\_recipient: Address of the new recipient)}{
    \If{msg.sender $==$ organization}{
        recipients[\_recipient] $\gets$ true\;
    }
}

\Fn{RemoveRecipient(\_recipient: Address of the recipient to be removed)}{
    \If{msg.sender $==$ organization}{
        recipients[\_recipient] $\gets$ false\;
    }
}

\Fn{RegisterBankAccount(\_recipient: Recipient address, \_account: Account to be registered)}{
    \If{msg.sender $==$ organization \textbf{and} recipients[\_recipient] $==$ true \textbf{and} bytes(\_account).length $>$ 0}{
        bankAccounts[\_recipient] $\gets$ keccak256(abi.encodePacked(\_account))\;
        \textbf{emit} BankAccountRegistered(\_recipient, bankAccounts[\_recipient])\;
    }
}

\caption{Authorization Functions}
\end{algorithm}
Upon deployment on the blockchain, the constructor function [Algorithm 1] starts the aNp contract and attributes the deploying address as the authoritative "organization," in charge of carrying out crucial operations and maintaining the beneficiary list. The functions AddRecipient and RemoveRecipient [Algorithm 1] give the organization the power to control the recipient list, and RegisterBankAccount [Algorithm 1] uses keccak256 encryption to securely link recipient addresses to bank accounts, with the organization controlling the registration process of these banking details. Each of these tasks helps to create a safe system with well defined roles and duties, and centralized bank account management increases system security.

The depicted algorithm [Algorithm 2] outlines a series of transaction functions pertinent to the operational mechanics of a blockchain-based financial distribution system. The functions [Algorithm 2] provide a methodological framework for the secure and verifiable management of balances and allowances. 

\begin{algorithm}
\LinesNotNumbered
\DontPrintSemicolon
\SetAlgoNlRelativeSize{-1}
\SetNlSty{}{}{}
\SetAlgoNlRelativeSize{0}
\SetKwInOut{KwData}{Input}
\SetKwInOut{KwResult}{Output}
\SetKwProg{Fn}{Function}{}{end}
\SetKw{KwRet}{Return}

\KwData{msg.sender: Current function caller, \_amt: Amount to be added}
\KwResult{Add funds to the organization's balance}
\If{msg.sender $==$ organization}{
    balances[organization] $+=$ \_amt\;
    \textbf{emit} FundsAdded(msg.value)\;
}

\KwData{msg.sender: Current function caller, \_recipient: Address of the recipient, \_amount: Amount to be sent}
\KwResult{Send allowance to a recipient}
\If{msg.sender $==$ organization \textbf{and} recipients[\_recipient] $==$ true \textbf{and} balances[organization] $\geq$ \_amount}{
    balances[organization] $-=$ \_amount\;
    \textbf{emit} AllowanceSent(\_recipient, \_amount)\;
}

\KwData{msg.sender: Current function caller}
\KwResult{Return the balance of the function caller}
\KwRet balances[msg.sender]\;

\caption{Transaction Functions}
\end{algorithm}

The balance of the organization is increased by the quantity specified by the parameter '\_amt' in the first function. This action requires that the'msg.sender' function be called by the organization itself. Following the effective addition of funds, the transaction's value is recorded in the event 'FundsAdded'. The'sendAllowance' function controls the distribution of allowances to recipients. 

This function [Algorithm 2] requires the following prerequisites to be met; the sender must be the organization, the recipient must be authorized, and the organization's balance must be greater than or equal to the quantity to be sent. If these conditions are met, the specified amount is deducted from the organization's balance, and the 'AllowanceSent' event is triggered, documenting the recipient and the amount transferred. 

The final function returns the caller's balance, encapsulating a fundamental aspect of the system's transparency. Collectively, these transaction functions [Algorithm 2] uphold the principles of secure, transparent, and accountable financial operations within the blockchain-enabled system, reflecting the robustness and precision inherent to smart contract-driven financial systems.

\subsubsection{Smart Contract Overview}
In the consortium blockchain environment of Hyperledger Besu, the smart contract facilitates both authorization and transactional activities. Technically, this contract utilizes a hierarchical access model, which centralizes control around an entity referred to as the organization. Such a design is essential for governing access rights and assuring transactional consistency across a decentralized architecture.

The requirement to ensure automation poses a significant challenge for this design. Each function, while designed to operate independently, must also integrate seamlessly into the larger system architecture. The persistent emphasis on the msg.sender attribute, which is essential for transactional accountability, introduces complex complexities. Such complexities necessitate stringent logic validation to prevent potential security vulnerabilities, particularly during operations with a high level of privilege.

Moreover, ensuring that the smart contract remains impervious to vulnerabilities while preserving operational efficiency presents a technical challenge. Given the specific requirements of financial systems on a consortium blockchain, this balancing act becomes even more complex. Conclusively, the smart contract exemplifies the complex challenges of fusing technical precision with the practical needs of financial systems in the context of a consortium blockchain.

\subsection{Hyperledger Besu Integration}
The open-source Ethereum client Hyperledger Besu functions seamlessly in both public and private environments. Besu typically employs consensus algorithms such as Ethash in public networks and anticipates Ethereum 2.0's Proof-of-Stake. For private networks, Besu incorporates consensus systems such as IBFT 2.0 and Clique Proof-of-Authority, targeting enhanced privacy and efficiency. The proposed system employs Hyperledger Besu with the IBFT 2.0 consensus mechanism \cite{nb25}, which operates with four validators, to assure network consensus even in the presence of potentially hostile nodes.

The IBFT 2.0 consensus mechanism functions within an environment interspersed with honest and Byzantine nodes. Byzantine nodes can act unpredictably, while honest nodes strictly follow the protocol. A pivotal feature in this system is the concept of a quorum. To ensure consensus, the protocol mandates the presence of a quorum, which is effectively represented by the relationship \( F(n) = \frac{N-1}{3} \) \cite{nb25}. 

This equation implies that for consensus to be achieved, a quorum, more than two-thirds of the participating nodes, must concur. So, while the system can tolerate up to \( F(n) \) Byzantine nodes, maintaining consensus requires \( n - F(n) > 2F(n) \) \cite{nb25}. Leveraging Ethereum's Ð{$\Xi$}Vp2p for message dissemination, IBFT 2.0 operates within an eventually synchronous network \cite{nb25}. A notable point in this network is the global stabilization time (GST), after which the delay in message transmission, denoted by \( \Delta \), becomes constant. Prior to GST, however, message delays can be unbounded, and message losses are plausible. This blend of Byzantine fault tolerance, the necessity for a quorum, and the nuances of synchronization form the crux of IBFT 2.0, ensuring its efficacy in achieving consensus across a diverse network \cite{nb25}.

Deploying a smart contract on Hyperledger Besu varies between public (\( P_{pub} \)) and private (\( P_{priv} \)) networks, despite both utilizing IBFT 2.0 PoA consensus. \( DP_{priv} \) entails deploying to a fixed set of validators \( V = 4 \), ensuring faster deployment time (\( DT_{priv} \)) and consistent consensus. On the other hand, \( DP_{pub} \) confronts a broader, dynamic validator set, potentially causing longer \( DT_{pub} \) durations. While the private deployment emphasizes restrictive access and enhanced control, the public one accommodates a more extensive range of participants. Thus, the deployment process and governance diverge substantially based on the network's nature.
\begin{algorithm}
\LinesNotNumbered
\DontPrintSemicolon
\SetAlgoNlRelativeSize{-1}
\SetNlSty{textbf}{(}{)}
\SetAlgoNlRelativeSize{0}
\SetKwInOut{Input}{Input}
\SetKwInOut{Output}{Output}
\SetKwProg{Fn}{Function}{}{end}
\SetKw{KwRet}{Return}

\SetKwBlock{KwProvider}{PrivateKeyProvider}{end}
\SetKwBlock{KwNetworks}{networks}{end}
\SetKwBlock{KwBesugo}{besugo}{end}

\Input{privateKeys, rpcUrl, min, max}
\Output{privateKeyProvider}

\KwProvider{
    privateKeyProvider $\gets$ new PrivateKeyProvider(privateKeys, rpcUrl, min, max) \;
    \KwRet privateKeyProvider \;
}

\KwNetworks{
    besugo \{
    \KwRet privateKeyProvider, network\_id: '1337', gas: 4500000, gasPrice: 0 \;
    \}
}

\Fn{Main}{
    besugo = networks.besugo \;
    \KwRet besugo \;
}

\caption{Configuring Hyperledger Besu Network}
\end{algorithm}

This algorithm [Algorithm 3] presents the configuration procedure for the private (IBFT 2.0)  Hyperledger Besu network in a structured and comprehensive manner. The algorithm commences by declaring four input parameters which are privateKeys, rpcUrl, min, and max. privateKeys refers to the keys required for network access, rpcUrl is the remote procedure call URL used by the system to communicate with the Hyperledger Besu node, min and max define the range within which the specific private key is selected. This algorithm's output is privateKeyProvider, a crucial component that enables secure network transactions. 

The PrivateKeyProvider block is the crux of the algorithm, which generates a secure connection with the blockchain node by creating a new instance of the PrivateKeyProvider with the declared inputs. The networks block subsequently establishes the configuration for the private Hyperledger Besu network. It includes the privateKeyProvider as a parameter and sets the network\_id to '1337', which is a common network identifier used in Ethereum-based networks for private local development. 

Moreover, the gas and gasPrice parameters are configured. Gas refers to the computational effort required to execute an operation, whereas gasPrice represents the price per unit of gas. Finally, the Hyperledger Besu network is derived from the previously configured networks in the Main function. Then, the Main function returns this Hyperledger Besu network, which is established with the correct parameters. This algorithm [Algorithm 3] is essentially a configuration guide for establishing a private (IBFT 2.0) local Hyperledger Besu network with the assistance of a Private Key Provider, thereby facilitating secure and efficient interactions with the blockchain.

The aNp smart contract is then introduced into this setting as the next phase. We assembled the Solidity-written contract and made it ready for deployment using the Truffle Suite, a well-known Ethereum development environment and testing framework. In this process, the configuration file for Truffle (truffle-config.js) is especially important. Here, we defined the required network specifications for redthe proposed system's private Hyperledger Besu network, such as the network ID, host, and port, as well as the "from" address for contract deployment.  
The Truffle Suite's migration script [Algorithm 4], another crucial component, is next developed. The procedure for deploying the smart contract to the network is automated by this script. 
\begin{algorithm}
\LinesNotNumbered
\DontPrintSemicolon
\SetAlgoNlRelativeSize{-1}
\SetNlSty{}{}{}
\SetAlgoNlRelativeSize{0}
\SetKwInOut{KwData}{Input}
\SetKwInOut{KwResult}{Output}
\SetKwProg{Fn}{Function}{}{end}
\SetKw{KwRet}{Return}

\KwData{aNp: Contract artifact for aNp}
\KwResult{Deploy the aNp contract}
\KwRet aNp.deployed()\;

\caption{Migration Script}
\end{algorithm}

Both the smart contract designated for deployment and the sequence of deployment are explicitly detailed. Once the migration script executes successfully, the smart contract undergoes deployment to the Hyperldger Besu network. Following this deployment, the smart contract's constructor process establishes the contract's state to a predefined initial value. Upon validation and integration into the blockchain, the contract engages with the proposed system's components, ensuring secure and efficient regulation of financial transactions.

This algorithm [Algorithm 4] concisely describes the deployment of the aNp contract artifact onto a blockchain network, namely the Hyperledger Besu network. This is accomplished by utilizing a migration script, a crucial component of Ethereum's Truffle framework. The algorithm [Algorithm 4] accepts the aNp contract artifact as input. In the context of Ethereum's smart contract, an artifact is a JSON file that contains crucial information about a contract, such as its opcode and the network addresses where it has been deployed. The aNp artifact represents the contract we discussed earlier that administers the organization's financial distribution.\newline
The algorithm's [Algorithm 4] output is the aNp contract that has been deployed, which is the consequence of calling the aNp.deployed() function. This function, which is part of the Truffle toolkit, returns a contract instance as it exists on the network. If a contract instance does not already exist, this function will create one. 

In conclusion, this algorithm [Algorithm 4] illustrates the application of Truffle for deploying a smart contract to a blockchain network. This is a crucial phase in the development of a blockchain application despite its apparent simplicity. It enables the contract's functionalities to be leveraged on the network, allowing the organization to conduct its financial distribution processes in a secure, transparent, and effective manner. Thus, the inclusion of Hyperledger Besu proved essential in creating a safe, open, and effective environment for the deployment of the smart contract.\newline
\begin{figure}[htp]
    \centering
    \includegraphics[width=.95\linewidth]{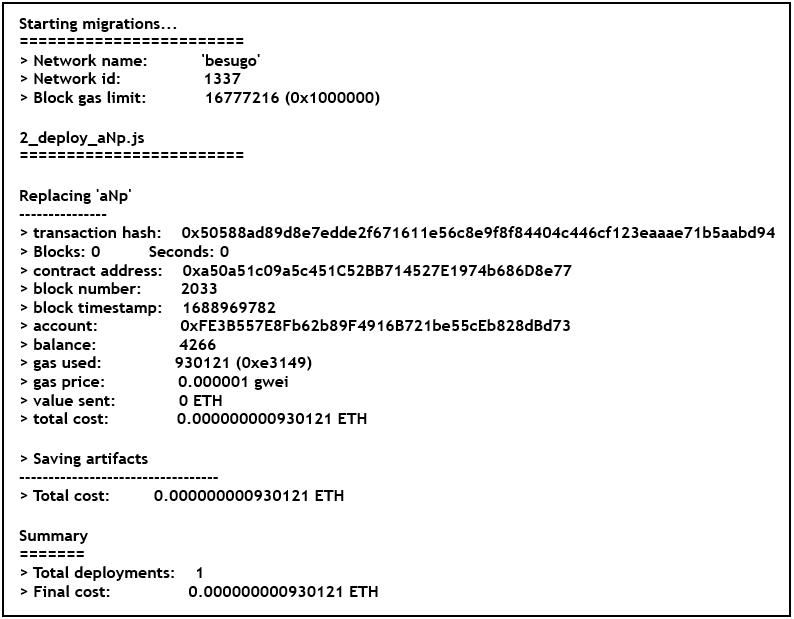}
    \caption{Smart Contract Deployment on HyperLedger Besu Private Network }
    \label{fig:Smart Contract Deployment on HyperLedger Besu Private Network  }
\end{figure}
The visualization [Fig. 4] presented in the conference paper depicts the successful deployment of the smart contract on the Hyperledger Besu network. This essential stage signifies the incorporation of the smart contract's functionality into the blockchain network, thereby enabling its secure and efficient operations. Transferring the contract's code onto the Besu network enables seamless interaction with the system's user interface during the deployment process. Consequently, this visualization [Fig. 4] serves as confirmation of the effective deployment and operational readiness of the smart contract within the private blockchain network, signifying the launch of a transparent, secure, and efficient financial distribution system.

\subsection{Proposed System's strength and limitation}
The proposed system, founded on a consortium blockchain utilizing Hyperledger Besu, represents a radical departure from the conventional financial distribution landscape. In response to the concerns expressed by previous research \cite{nb15,nb16,nb17,nb18,nb19,nb20,nb21,nb22,nb23,nb24} this paper introduces a sophisticated solution that leverages the capabilities of Hyperledger Besu, particularly its private configuration and the incorporation of the IBFT 2.0 consensus mechanism. The system employs the deterministic properties of smart contracts on the Hyperledger Besu platform to mitigate market risks, thereby ensuring consistent adherence to predefined terms and reducing the likelihood of market anomalies. 

Regarding data access controls, the private nature of Hyperledger Besu with IBFT 2.0 establishes an exclusive environment where data access remains restricted to authenticated participants, thereby preserving the integrity and confidentiality of data. While the immutable nature of the blockchain presents difficulties in data deletion, Hyperledger Besu's architecture enables effective data recovery and, via specific smart contract functionalities, ensures that data remains inaccessible or obsolete when required. Lastly, the proposed system is a robust consortium blockchain-based platform that optimally integrates Hyperledger Besu and smart contracts to resolve the identified gaps, thereby fostering advancements in the financial distribution domain.

Incorporating Hyperledger Besu into the system has significantly improved its functionality and security, but there are still some limitations. Primarily, the system's reliance on a permissioned network may hinder its scalability potential, particularly in terms of widespread adoption across a variety of financial institutions. The use of the IBFT 2.0 consensus mechanism, despite being effective at mitigating certain security concerns, may induce latency in transaction validations, particularly when transaction volumes are high. 

Moreover, to increasing automation and decreasing human intervention, smart contracts introduce the danger of irreversibility. Any defect or vulnerability in the contract logic can potentially lead to significant financial discrepancies, and rectifying such an error within a blockchain environment can be challenging. Lastly, the system's widespread adoption is contingent on the familiarity and trust of financial institutions in Hyperledger Besu, which may differ by region and stakeholder.

\section{Conclusion}
The combination of Hyperledger Besu and smart contract technology paves the way for the revolutionary transformation of financial systems. Through the investigation and implementation of the aNp smart contract within a private Hyperledger Besu network, we have demonstrated a secure and efficient model for managing financial transactions, thereby highlighting the applicability of this approach in a variety of financial contexts. Future prospects for such systems are indeed promising, ranging from enhancing fraud detection and regulatory compliance to enhancing supply chain finance, international trade, and cross-border payments efficiencies. When combined with other emergent technologies such as AI and IoT, the potential for additional innovation and transformation of financial systems becomes nearly limitless. Nevertheless, it is essential to emphasize the need for additional research and development to resolve potential scalability, performance, and regulatory issues associated with blockchain-based solutions. Nevertheless, the proposed aNp smart contract system is a valuable starting point and an inspiring demonstration of the potential of blockchain technology, specifically Hyperledger Besu, in financial systems. The convergence of finance and technology will continue to redefine the comprehension of financial systems and processes as this field of study continues to develop.

\end{document}